\documentclass[prb,twocolumn,showpacs]{revtex4-1}
\usepackage{graphicx}
\usepackage{amsmath}
\usepackage{amssymb}

\renewcommand{\Im}{\mathrm{Im}\,}

\renewcommand{\Im}{\mathrm{Im}\,}

\newcommand{\ra}{\rangle}
\newcommand{\la}{\langle}
\DeclareMathAlphabet{\bi}{OML}{cmm}{b}{it}
\begin{document}
\title{Modulation effect on the spin Hall resonance}

\author{SK Firoz Islam}
\affiliation{Department of Physics, Indian Institute of Technology-Kanpur,
Kanpur-208 016, India}

\begin{abstract}
The effect of a weak electrical modulation on spin Hall resonance is
presented here. In presence of the magnetic field normal to the plane
 of the motion of electron, the Landau levels are formed which get
 broadened due to the weak modulation. The width of the Landau levels 
broadening are periodic with the inverse magnetic field. There is a certain 
magnetic field for which the crossing of Landau levels between spin-up and
 spin-down branches takes place. This gives rise to the resonance in the spin Hall conductivity (SHC).
 The Landau levels broadening or the energy correction due
 to the modulation removes the singularity appears at the resonance field in SHC,
leading to the suppression of SHC accompanied by two new peaks around this point.
 The separation of these two peaks increases with the increase of the modulation
 period. Moreover, we find that the height of the two peaks are also modulation period dependent.
\end{abstract}
\date{\today}


\maketitle
\section{Introduction}
The lack of structural symmetry in a two dimensional electron gas 
(2DEG) has attracted lots of  attention
of condensed matter physicists due to the possible remarkable future 
applications as proposed by Datta and Das\cite{das1,appl1,appl2,appl3}. 
Various experiments have confirmed the existence of a spin-orbit
coupling known as Rashba spin-orbit coupling (RSOC)
in a 2DEG with structural inverse asymmetry\cite{dutta}.
The RSOC lifts the spin degeneracy of electron even in absence
 of any magnetic field and provide a nice way to 
manipulate electron's spin degree of freedom. Moreover, the RSOC is responsible for
 some noble effects like spin-FET\cite{das1,bandyo}, metal-insulator 
transition in a two-dimensional hole gas\cite{mit}, 
spin-resolved ballistic transport\cite{ballistic} and spin-galvanic effect\cite{galvanic}. The
experimental confirmation of the enhancement of RSOC strength up 
to a significant amount by the application of the gate voltage has made this field
much more attractive\cite{tech,matsu}.
\\
The RSOC produces two different energy branches for spin-up and spin-down electrons
\cite{rashba,rashba1} and manifests itself through transport\cite{dutta,wang,wang1},
 magnetic\cite{zhang} and
optical properties \cite{optical}. The spin Hall effect (SHE)\cite{perel,she} is
one of the most unique property exhibits by 2DEG with RSOC. The SHE has
been studied extensively theoretically after the experimental detection
of the spin accumulation on opposite sides of the sample with opposite
polarization\cite{kato,sinova}.
\\
In presence of magnetic field, an additional degeneracy comes into picture when
Landau levels of two opposite branches overlap for a certain magnetic field.
This degeneracy is the outcome of the competition between Zeeman splitting and Rashba splitting.
This additional degeneracy manifests through a jump in the
spin polarization and the resonance in SHC\cite{shen}. 
\\
Later, the effect of disorder in the spin Hall resonance has been studied
theoretically and showed the appearance of double peaks around the resonance field 
instead of single peak at the resonance point\cite{disorder}. The splitting of single peak into double
is the effect of the disorder induced Landau levels broadening.
\\
The  disorder is not the only sole way to broaden the width of Landau levels.
The Landau levels can be broadened by applying a weak electric or magnetic modulation\cite{peeters1,peeters2}.
The width of the Landau levels due to the modulation is periodic with inverse
magnetic field whereas the disorder induced broadening is magnetic field independent. This
kind of modulation induced oscillatory broadening of Landau levels results in 
oscillatory diffusive conductivity in inverse magnetic field, which is known 
as Weiss oscillation\cite{peeters1,peeters2}. Moreover, the modulation
effect on the spin Hall conductivity has been studied theoretically also
and shows the Weiss type oscillation accompanied by beating in SHC at low magnetic field\cite{wang2}.
 In this paper, our major aim is to study the effect of
modulated Landau levels in spin Hall resonance.

This paper is organized as follows. In section II, we summarize the 
exact energy eigenvalues and the corresponding eigenfunctions of a 2DEG 
with the Rashba SOI in presence of a perpendicular uniform magnetic field.
The formalism is given in section III for the calculation of SHC.
Numerical results are are presented in section IV. 
We present a summary of our work in Sec. V.
\section{Energy spectrum and energy correction due to the modulation in presence of the Rashba SOI}
The Hamiltonian of an electron $(-e)$ with the RSOC 
in presence of a perpendicular magnetic field ${\bf B} = B \hat z$ is given by
\begin{equation}
H = \frac{({\bf p} + e {\bf A})^2}{2m^{\ast }}\sigma_0 + 
\frac{\alpha}{\hbar }\left[ {\mbox{\boldmath $\sigma$} } \times 
({\bf p}+e{\bf A})\right]_z + \frac{1}{2}g\mu _{_B} B \sigma_z, \label{Ham}
\end{equation}
where ${\bf p}$ is the two dimensional momentum operator, $m^{\ast }$ is the effective mass 
of the electron, $g$ is the Lande-g factor, $ \mu_{_B} $ is the Bohr magneton,
$\sigma_0 $ is the unit matrix, 
${\mbox{\boldmath $\sigma$} }=(\sigma_x,\sigma_y,\sigma_z)$ are the Pauli 
spin matrices, and $\alpha $ is the strength of the RSOC.

Here, we shall just mention the exact solutions of 
the Hamiltonian $H$. Using the Landau wave functions without 
the RSOC as the basis, one can obtain the energy 
spectrum and the corresponding eigenfunctions\cite{wang}. 
For $n=0$  there is only one level, the same as the lowest Landau
level without RSOC, with energy
$ E_0^+ = E_0 = (\hbar \omega - g \mu_{_B} B)/2 $
and the corresponding wave function is
\begin{equation}
\Psi_0^+ (k_y)= \frac{e^{i k_y y}}{\sqrt{L_y}} \phi_0(x + x_0)
\left(
\begin{array}{c}
0 \\
1
\end{array}
\right).
\end{equation}
Here, $ \omega = eB/m^* $, $x_0 = k_y l_0^2$ with 
$l_0 = \sqrt{\hbar/(eB)}$ is the magnetic length scale.
For $n=1,2,3....$ there are two branches of the energy levels, 
denoted by $+$ corresponding to the "spin-up" electrons 
and $-$ corresponding to the "spin-down" electrons with energies
\begin{equation}
E_{ns} = n\hbar\omega+{s} \sqrt{E_0^2 + n E_{\alpha} \hbar \omega},
\end{equation}
where $s=\pm$ and $ E_{\alpha} = 2 m^{\ast} \alpha^2/\hbar^2$ is the Rashba energy determined
by the SOI strength $ \alpha $.
The corresponding wave function for $+$ branch is
\begin{equation}
\Psi _n^+ (k_y)=\frac{e^{i k_y y}}{\sqrt{L_y A_n }}\left(
\begin{array}{r}
D_n \phi_{n-1}(x + x_0)
\\
\phi_n (x + x_0)
\end{array}
\right) \text{,}
\end{equation}
and the $-$ branch is
\begin{equation}
\Psi_n^-(k_y)=\frac{e^{i k_y y}}{\sqrt{L_y A_n }}\left(
\begin{array}{r}
\phi_{n-1}(x + x_0)
\\
- D_n \phi_n(x + x_0)
\end{array}
\right) \text{,}
\end{equation}\\
where $A_n = 1 + D_n^2 $ and
$ D_n = \sqrt{n E_{\alpha} \hbar \omega}/
[E_0 + \sqrt{E_0^2 + n E_{\alpha} \hbar \omega}]$.
Here, $\phi_n(x) = (1/\sqrt{\sqrt{\pi }2^n n!l_0}) 
e^{-x^2/2l_0^2} H_n(x/l_0) $ is the normalized harmonic 
oscillator wave function with $n$ is the Landau level index,
\\
The condition of the crossing of the energy levels between
$E^{+}_{n} $ and $E^{-}_{n+1}$ can be obtained from
$E^{+}_{n}-E^{-}_{n+1}=0$, which gives\cite{shen}
\begin{equation}
\sqrt{(1-g^*)^2+\eta n}+\sqrt{(1-g^*)^2+\eta (n+1)}=2,
\end{equation}
where $\eta=4 E_{\alpha}/(\hbar\omega)$, $g^*=gm^*/(2m_0)$ and the quantum number $n$
follows $2n\le \nu\le 2(n+1)$ with $\nu=N_e/N_{\phi}$ being the filling fraction. Here,
$N_\phi$ is the magnetic flux quanta. The magnetic field
at which the above condition satisfy is termed as the resonance point.
\\
\\
Now  a weak spatial electric modulation $V=V_0\cos(Kx)$ is applied to the system along the
$x$-direction with periodicity $a$, where $K=2\pi/a$.
The first-order energy correction due to this modulation can be easily derived\cite{wang1} 
as $\Delta E_{n,k_y,s}=\frac{V_u}{A_n}F_{n,s} $, where $V_u=V_0\exp(-u/2)\cos(Kx_0)$
with $u=K^2l_0^2/2$, $ F_{n,+}(u)=D^2_{n}L_{n-1}(u)+L_{n}(u)$ and $F_{n,-}(u)=D^2_{n}L_{n}(u)+L_{n-1}(u)$
. Here, $L_n(u)$ is the Laguerre polynomial of order $n$.
\section{Formalism}
We considers a Rashba coupled 2DEG in a $x-y$ plane with size $L_x\times L_y$. In presence of a weak 
electric field along the $x$-direction, the electron with opposite spin  scattered
 towards $y$-axis of the sample and gives rise to SHC. To calculate the SHC, we shall use the Kubo formula as
\begin{eqnarray}\label{shc}
 \sigma^{\rm SHC}_{xy}&=&-\frac{e\hbar}{L_xL_y}\sum_{\{n,s\}\neq{\{n',s'\}},k_y}[f(E_{n,s})-f(E_{n's'})]\nonumber\\
&\times&\frac{\Im\la n,k,s\mid J_y^{S_z}\mid n',k,s'\ra
\la n',k,s'\mid v_x \mid n,k,s\ra}{(E_{ns}-E_{n's'})^2+\delta^2}\nonumber\\
\end{eqnarray}
with the $z$-component of spin current operator along $y$-direction $J_y^{S_z}=\frac{1}{2}\{v_y,S_z\}$
as per the conventional definition. Here, $\delta$ is assigned to control the
 hight of the resonance peak. The velocity operators are
\begin{equation}
 v_x=\frac{p_x}{m^*}\sigma_0-\frac{\alpha}{\hbar}\sigma_y,
\end{equation}
 \begin{equation}
 v_y=(x+x_c)\omega_c\sigma_0+\frac{\alpha}{\hbar}\sigma_z,
\end{equation}
and $f(E_{n,s})$ is the equilibrium Fermi distribution function. As per the 
restriction imposed over summation in the expression of SHC there is no intra-band $(\{n,s\}=\{n',s'\})$
contribution.
\\
For $n'=n+1$, the matrix elements of spin current operator and velocity operator are
\begin{eqnarray}
 \la n,k,+\mid J_y^{S_z}\mid n+1,k,-\ra&=&\frac{1}{\sqrt{A_nA_{n+1}}}\frac{\hbar\omega}{2}\frac{l_0}{\sqrt{2}}\nonumber\\
&\times&[D_n\sqrt{n}+D_{n+1}\sqrt{n+1}]\nonumber\\
\end{eqnarray}
and
\begin{eqnarray}
 \la n+1,k,-\mid v_x\mid n,k,+\ra&=&i\frac{1}{\sqrt{A_nA_{n+1}}}\frac{\hbar}{\sqrt{2}m^*l_0}\nonumber\\
&\times&[D_n\sqrt{n}-D_{n+1}\sqrt{n+1}+\sqrt{2}\zeta],\nonumber\\
\end{eqnarray}
where $\zeta=\sqrt{2}\frac{\alpha m^* l_0}{\hbar^2}$.
In presence of modulation the SHC expression need to be modified
 by including the first order energy correction as $E^m_{ns}\backsimeq E_{n,s}+\Delta E_{n,k_y,s}$.
Now substituting the above matrix elements in Eq. (\ref{shc}) the SHC of modulated system is given by
\begin{eqnarray}
 \sigma^{\rm SHC}_{xy}&=&-\frac{e}{8\pi}\frac{l_0^2}{a}\int_{0}^{a/l_0^2}dk_y\sum_{n}[f(E^m_{n,+})-f(E^m_{n+1,-})]\nonumber\\
&\times&\frac{\lambda_{n,\zeta}}
{[(E^m_{n,k,+}-E^m_{n+1,k,-})/\hbar\omega]^2}
\end{eqnarray}
whereas
\begin{eqnarray}
\lambda_{n,\zeta}&=&\frac{1}{A_nA_{n+1}}[D_n^2n-D^2_{n+1}(n+1)+\sqrt{2}\zeta(D_n\sqrt{n}\nonumber\\&+&D_{n+1}\sqrt{n+1})].
\end{eqnarray}
Here, the modulation induced broadening will be acting to reduce the possibility of the
occurrence of resonance.
\section{results and analysis}
For the numerical calculation, we use the following parameters: 
electron effective mass $m^*=0.05m_e$ with $m_e$ as the free electron 
mass, RSOC strength $\alpha=0.9\alpha_0$ with $\alpha_0=10^{-11}$ eV-m and the
carrier concentration $n_e=1.9\times10^{16}/m^2$. These parameters are same as used 
in several theoretical works for which the resonance appears around $B\simeq 6.1$ T.
\\
We plot the SHC with inverse magnetic field around the resonance
 point for different modulation periods in Fig.1.
\begin{figure}[t]
\begin{center}\leavevmode
\includegraphics[width=98mm]{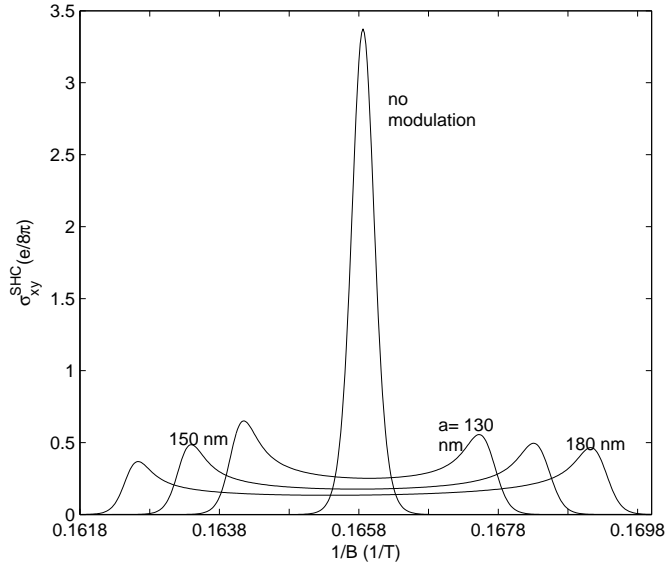}
\caption{Plots of the SHC vs  inverse magnetic field $B$ for different
modulation period. Modulation strength $V_0=3$ meV,
the temperature $T=0.6$ K}
\label{Fig1}
\end{center}
\end{figure}
It shows that the inclusion of modulation splits the resonance peak into two weak peaks
symmetrically on both sides of the resonance point. The physical origin of the appearance
of double peaks due to the Landau levels broadening
has been discussed in Ref\cite{disorder}. However, in our case
the Landau level's broadening is due to the modulation instead of disorder.
The resonance point now shows a minimum in conductivity and maintains 
it for a wide range of magnetic field. The gap between two new peaks which appears
on both sides of the resonance field are modulation period dependent.
The gap is increasing with increasing modulation period. Note that two new peaks
are not confined within the central peak which was observed in the 
disorder induced Landau broadening case\cite{disorder}.
Another point is that the intensity of the two new peaks are
decreasing with the increase of the modulation period which can
be explained from the expression of energy correction as follows:
 $\Delta E_{n,k_y,s} \propto V_u=V_0\exp(-u/2)=V_0\exp\{-(\pi l_0/a)^2\}$ i.e; the
increasing modulation period enhances the effective modulation strength.

\begin{figure}[t]
\begin{center}\leavevmode
\includegraphics[width=98mm]{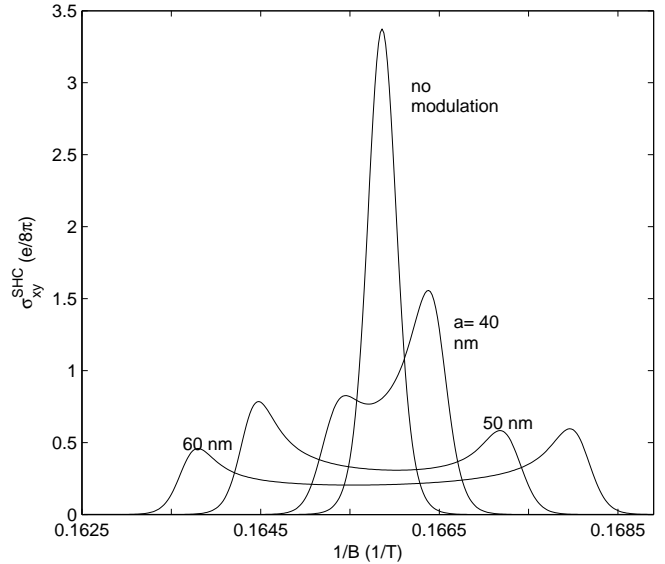}
\caption{Plots of the SHC vs  inverse magnetic field $B$ for short range
modulation period.  }
\label{Fig1}
\end{center}
\end{figure}
In Fig.2, we plot the same but for shorter range of modulation periods.
This shows the disappearing of double peaks with decreasing the modulation periods.
It is beacause of the reduction of the effective modulation strength with decreasing
periods.  Finally, for sufficiently low periodicity two peaks collapsed to a single peak.
\begin{figure}[t]
\begin{center}\leavevmode
\includegraphics[width=98mm]{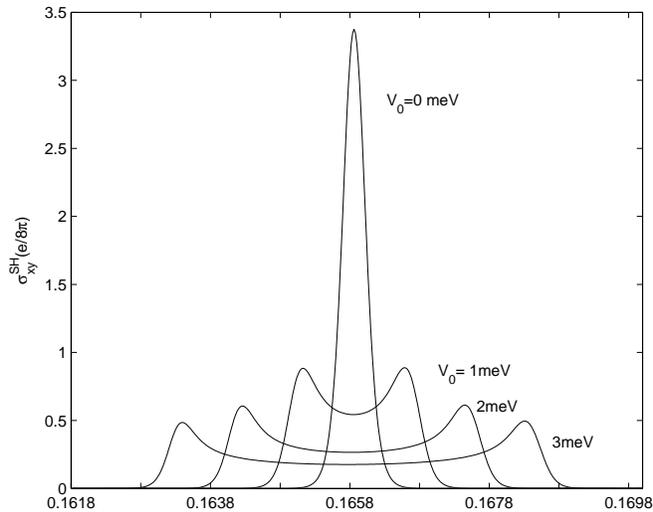}
\caption{Plots of the SHC vs  different strength of the modulation for $a=500$ nm.}
\label{Fig1}
\end{center}
\end{figure}
Fig.3 shows the SHC response for different strength of modulation.
The increasing of modulation strength results in suppressing the SHC more.
At the same time the flat band between two weak peaks also increases with
the increase of the modulation strength. The SHC at the resonance point is 
decreasing with the increase of the strength of modulation.
 \\
 In our case, the Landau levels
broadening is oscillatory with modulation periods as well as with inverse magnetic field. At
Fermi energy, it's  oscillation period\cite{firoz} is $2\hbar(k_F\mp k_{\alpha})/(eaB)$ with $k_F$
the Fermi vector and $k_{\alpha}=m^*\alpha^2/\hbar^2$. The level width can be tuned by
changing $a$. This width modulation can not be captured in the small range of 
magnetic field which is actually taken here around the resonance point. The changing periods
results itself only through the suppressing or enhancing the intensity of the double peaks
and by widening or shortening the gap between these peaks.
\section{summary}
We study the effect of modulation induced Landau levels
broadening in spin Hall resonance. We find that this level broadening
suppress the resonance SHC and
splits the central resonance peak into double peaks but relatively weak in
strength. The separation between two new peaks can be controlled by the modulation period
i.e; the gap increases with increasing modulation period. Moreover,
with shortening the modulation periods the intensity of the two new peaks starts two differ.
A wide flat band is observed between two peaks in our case whereas in 
disorder induced level broadening there is a narrow region between two peaks.
The increasing strength of modulation reduces the intensity of double peaks and
widens the flat band between two peaks.

\section{Acknowledgement}
I would like to thanks to Dr. T. K. Ghosh for usefull discussions.
 This work is financially supported by the CSIR, Govt.of India under the grant
CSIR-SRF-09/092(0687) 2009/EMR F-O746.

\end{document}